\documentclass[twocolumn,showpacs,preprintnumbers,amsmath,amssymb,prl]{revtex4}

\usepackage{graphicx}
\usepackage{dcolumn}

\usepackage{mathptmx, courier, pifont}
\usepackage[scaled=0.92]{helvet}
\usepackage[T1]{fontenc}
\usepackage{textcomp}

\usepackage{bm}

\begin{document}

\title{$\gamma$-vibrational states in superheavy nuclei}

\author{Yang Sun$^{1,2,3}$, Gui-Lu Long$^{4}$, Falih Al-Khudair$^{4}$,
Javid A. Sheikh$^{5}$}

\affiliation{ $^{1}$Department of Physics, Shanghai Jiao Tong
University, Shanghai 200240, People's Republic of China \\
$^{2}$Institute of Modern Physics, Chinese Academy of Sciences,
Lanzhou 730000, People's Republic of China \\
$^{3}$Joint Institute for Nuclear
Astrophysics, University of Notre Dame, Notre Dame, IN 46556, USA \\
$^{4}$Department of Physics, Tsinghua
University, Beijing 100084, People's Republic of China \\
$^{5}$Department of Physics, University of Kashmir, Hazrathbal,
Srinagar, Kashmir 190 006, India}

\date{\today}

\begin{abstract}
Recent experimental advances have made it possible to study excited
structure in superheavy nuclei. The observed states have often been
interpreted as quasi-particle excitations. We show that in
superheavy nuclei collective vibrations systematically appear as
low-energy excitation modes. By using the microscopic Triaxial
Projected Shell Model, we make a detailed prediction on
$\gamma$-vibrational states and their $E2$ transition probabilities
to the ground state band in Fermium and Nobelium isotopes where
active structure research is going on, and in $^{270}$Ds, the
heaviest isotope where decay data have been obtained for the
ground-state and for an isomeric state.
\end{abstract}

\pacs{21.10.Re, 21.60.Cs, 23.20.Lv, 27.90.+b}

\maketitle


One of the important predictions in nuclear physics is the emergence
of a region of long-lived superheavy elements beyond the actinides,
the so-called `island of stability'. The question concerns the
precise location of the next closed nucleon shells beyond $Z=82$ and
$N=126$. To reach the island, much of the experimental effort has
been focused on the direct creation of superheavy elements. In
recent years, progress has also been made in structure studies for
nuclei beyond Fermium, thanks to the development of detector systems
for decay and in-beam studies using recoil separators and heavy ion
fusion reactions \cite{Leino04,Rodi04}. It has been suggested that
by studying the transfermium nuclei, in particular their excited
structure, one can gain useful information on relevant
single-particle states \cite{Chas77}, which is key to locating the
island.

The nuclei of our interest, the Fm ($Z=100$) and No ($Z=102$)
isotopes, belong to the heaviest mass region where structure can
currently be studied experimentally. The yrast property of these
nuclei shows that they are generally good rotors. Rotational
behavior of some of these yrast bands has been successfully
reproduced by several models (see, for example, Refs.
\cite{ER00,Dug01,Ben03,Afa03}). The discussion on excited
configurations so far have been focused on quasi-particle
excitations \cite{Rodi07,Rodi06,Tandel06,Green07}. On the other
hand, a deformed rotor can, according to the collective model,
undergo dynamical oscillations around the equilibrium shape,
resulting in various low-lying collective vibrational states.
Ellipsoidal oscillation of the shape is well known as $\gamma$
vibration \cite{Bohr52}. It is thus natural to consider $\gamma$
vibrational states in superheavy nuclei, and in fact, this
excitation mode has begun to draw one's attention
\cite{Hess03,Gogny06}. Knowledge on vibrational states in superheavy
nuclei is particularly useful for this less known mass region
because of the interpretation of the observed low-lying
spectroscopy.

Early calculation of $\gamma$-vibrational states in heavy nuclei was
performed by Marshalek and Rasmussen \cite{MR63} using the
qausi-boson approximation, and by B\`es {\it et al.} \cite{Bes65}
using the quadrupole-plus-pairing model based on deformed Nilsson
states \cite{Nilsson55}. Modern treatment of $\gamma$ vibration
includes the Tamm-Dancoff method, the random phase approximation
\cite{Egido80}, and others \cite{RS80,So92}. In Ref. \cite{Sun00}, a
shell-model-type method for describing $\gamma$-vibrational states
was introduced, which is based on the Triaxial Projected Shell Model
\cite{SH99}, an generalized version of the original Projected Shell
Model \cite{PSM} by extending it to a triaxially deformed basis. It
was shown \cite{Sun00} that by performing diagonalization in a basis
constructed with exact three-dimensional angular-momentum-projection
on triaxially deformed states, it is feasible to describe
$\gamma$-vibrational states in a shell model framework. In this way,
one can achieve a unified treatment of ground-state band ($g$ band)
and multiphonon $\gamma$-vibrational bands ($\gamma$ band) in one
calculation, and the results can be quantitatively compared with
data. The underlying physical picture of generating
$\gamma$-vibration in deformed nuclei is analogous to the classical
picture of Davydov and Filippov \cite{DF58}. Subsequent papers
\cite{Sheikh01,Sun02,Bout02} studied electromagnetic transitions by
using wave functions obtained from the shell model diagonalization.

The Projected Shell Model \cite{PSM} is a shell model that uses
deformed bases and the projection technique. In the present
calculation, we use the triaxially-deformed Nilsson plus BCS basis
$\left| \Phi \right\rangle $. The Nilsson potential is
\begin{equation}
\hat{H}_0-\frac 23\hbar \omega \left[ \epsilon \hat{Q}_0+\epsilon
^{\prime}\frac{\hat{Q}_{+2}+\hat{Q}_{-2}}{\sqrt{2}}\right],
\label{Nils}
\end{equation}
where $\hat{H}_0$ is the spherical single-particle Hamiltonian
with inclusion of appropriate spin-orbit forces parameterized by
Bengtsson and Ragnarsson \cite{BR85}. The axial and triaxial parts
of the Nilsson potential in Eq. (\ref{Nils}) contain the
deformation parameters $\epsilon$ and $\epsilon ^{\prime}$
respectively, which are related to the conventional triaxiality
parameter by $\gamma = \tan^{-1}(\epsilon ^{\prime} / \epsilon )$.
Pairing correlation in the Nilsson states is taken into account by
a standard BCS calculation, within a model space of three major
shells for each kind of nucleon ($N$ = 5, 6, 7 for neutrons and
$N$ = 4, 5, 6 for protons).

The rotational invariant two-body Hamiltonian
\begin{equation}
\hat{H}=\hat{H}_0-\frac \chi2 \sum_\mu \hat{Q}_\mu ^{+}\hat{Q}_\mu
-G_M\hat{P}^{+}\hat{P}-G_Q\sum_\mu \hat{P}_\mu ^{+}\hat{P}_\mu
\label{Hamilt}
\end{equation}
is diagonalized in the three-dimensional
angular-momentum-projected basis $\left\{ \hat{P}_{MK}^I\left|
\Phi \right\rangle ,0\leq K\leq I\right\}$, where $\hat{P}_{MK}^I$
is the angular-momentum-projector \cite{RS80}. The wave function
thus takes the form
\begin{equation}
\left| \Psi^\sigma_{IM}\right\rangle =\sum_{0\leq K\leq
I}f^\sigma_{IK} \hat{P}_{MK}^I\left| \Phi \right\rangle ,
\label{Equat}
\end{equation}
where $\sigma$ specifies the states with the same angular momentum
$I$. The wave function (\ref{Equat}) is explicitly written as a
superposition of projected $K$ states. The two-body forces in Eq.
(\ref{Hamilt}) are quadrupole-quadrupole ($QQ$), monopole-pairing,
and quadrupole-pairing interaction, respectively. The strengths of
the monopole and quadrupole pairing forces are given respectively by
$G_M$ and $G_Q$ in Eq. (\ref{Hamilt}), where
\begin{equation}
G_M={{ G_1\pm G_2\frac{N-Z}A}\over A},
\label{Pairing}
\end{equation}
with $"+"$ for protons and $"-"$ for neutrons. We use $G_1=21.24$,
$G_2=13.86$, and assume the quadrupole-pairing strength in Eq.
(\ref{Hamilt}) to be $G_Q=0.13G_M$, which are found to be
appropriate for this mass region \cite{Rodi06,Falih07}. The
$QQ$-force strength $\chi$ is determined such that it holds a
self-consistent relation with the quadrupole deformation $\epsilon $
\cite{PSM}.

\begin{figure}
\includegraphics[width=7cm]{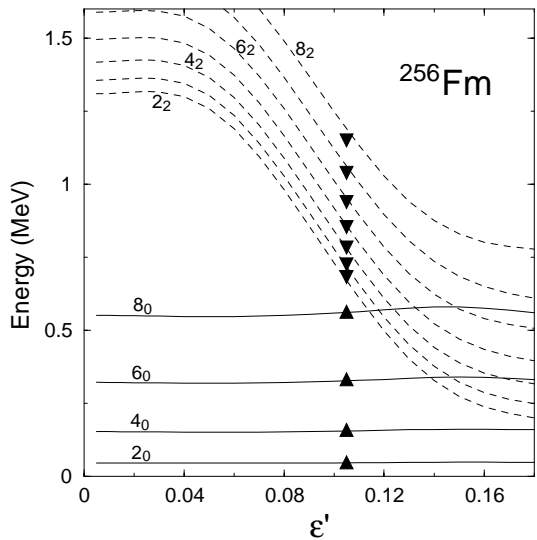}
\caption{Calculated energies of the $g$ band (solid lines) and
$\gamma$ band (dashed lines) in $^{256}$Fm as functions of
triaxiality parameter $\epsilon ^{\prime}$ for various angular
momenta. Each curve is labeled by $I_K$. The experimental $g$ band
(filled triangle-up) and possible $\gamma$ band (filled
triangle-down) are best reproduced with $\epsilon
^{\prime}=0.105$. Data are taken from Ref. \protect\cite{Fm256}.}
\label{fig1}
\end{figure}

The nucleus $^{256}$Fm is taken as our discussion example because it
is perhaps the only nucleus in this mass region where some observed
states were thought to be $\gamma$ vibrational states \cite{Fm256}.
Figure 1 shows the calculated energies as functions of triaxiality
parameter $\epsilon ^{\prime}$ for various angular momenta. Although
this figure looks similar to the one shown in the seminal paper of
Davydov and Filippov \cite{DF58}, it is obtained from a fully
microscopic theory. Unlike the asymmetric rotor model, our spectrum
depends not only on the deformation parameters but also
microscopically on the detailed shell filling. In Fig. 1, each curve
is labeled by $I_\sigma$. It turns out (see later discussions) that
for the nuclei studied in the present paper, $K$ mixing is very
small. Therefore, $\sigma\approx K$, and we can practically use
$I_K$ to label the states. One sees that, for the $g$ band ($K=0$)
in $^{256}$Fm, the energies are nearly flat as $\epsilon ^{\prime}$
varies, and the values remain very close to those at zero
triaxiality. Thus, we can conclude that the triaxial basis has no
significant effect on the $g$ band in such a superheavy rotor, and
one can practically describe $g$ band using an axially deformed
basis.

On the other hand, it has a large effect on excited bands with $K\ne
0$ (second and higher excited bands are not shown in this figure,
but will be discussed in Fig. 3). Their excitation energies are
indeed very high at $\epsilon ^{\prime}\approx 0$, but come down
quickly as the triaxiality in the basis increases. At $\epsilon
^{\prime}=0.105$, the first excited band ($K=2$) nicely reproduces
the observed $2^+$ band in $^{256}$Fm \cite{Fm256}. In our
calculation, the triaxial parameter $\epsilon ^{\prime}$ serves as a
free parameter adjusted to reproduce the $2^+$ bandhead. It should
be noted that the excited bands studied in this paper are obtained
by introducing triaxiality in the basis (quasiparticle vacuum). They
are {\it collective} excitations, not quasi-particle excitations.

The above results can be further understood by studying the $K$
mixing coefficients for each projected $K$ state in the total wave
function of Eq. (\ref{Equat}). It is found that for $^{256}$Fm (and
for all nuclei studied in this paper), $K$ mixing is negligibly
small. States in the $g$ band are essentially the projected $K=0$
state for any $\epsilon ^{\prime}$. The basis triaxiality does not
influence the $g$ band result, and the rotational behavior of the
$g$ band can be understood by using a simple axially-deformed rotor.
The excited bands are also built by rather pure projected $K$
states. For example, the first excited band with the bandhead spin
$I=2$ is mainly the projected $K=2$ state (labeled as $I_2$). The
fact of very small $K$ mixing in these nuclei sets up a favorable
condition for the occurrence of $K$ isomeric states
\cite{Rodi06,Tandel06,Xu04}.

\begin{figure*}
\includegraphics[width=13.5cm]{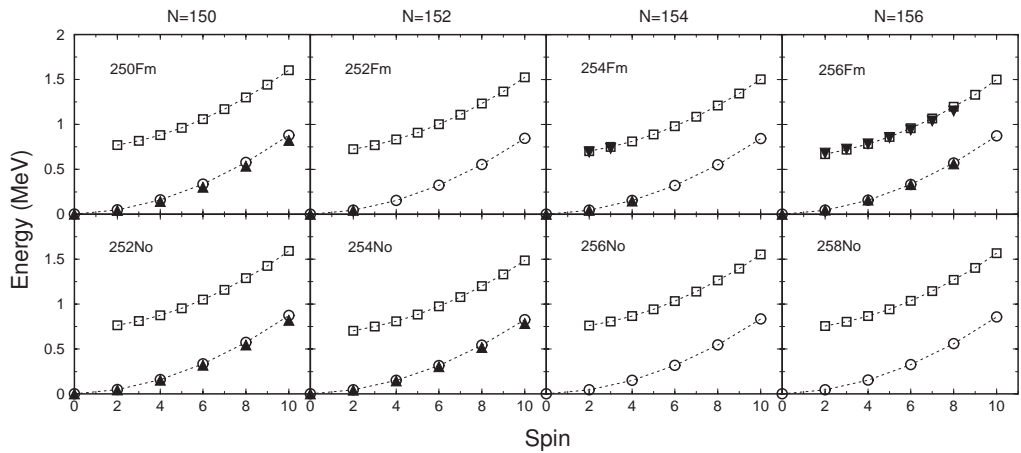}
\caption{Calculated energies for $g$ bands (open circles) and
$\gamma$ bands (open squares). Available experimental data (filled
triangle-up for $g$ bands and triangle-down for $\gamma$ bands)
are also shown for comparison. Data are taken from Refs.
\protect\cite{Fm250} ($^{250}$Fm), \protect\cite{Fm252}
($^{252}$Fm), \protect\cite{Fm254} ($^{254}$Fm),
\protect\cite{Fm256} ($^{256}$Fm), \protect\cite{No252}
($^{252}$No), and \protect\cite{No254} ($^{254}$No).} \label{fig2}
\end{figure*}

\begin{table}
\caption{Axial and triaxial deformation parameters used in the
calculation.} \label{tab:1}
\begin{tabular}{ccccccccc}
\hline\noalign{\smallskip}
 & $^{250}$Fm & $^{252}$Fm & $^{254}$Fm & $^{256}$Fm & $^{252}$No & $^{254}$No & $^{256}$No & $^{258}$No \\
\hline\noalign{\smallskip}

$\epsilon$ & 0.240 & 0.255 & 0.250 & 0.230 & 0.240 & 0.260 & 0.250 & 0.230 \\
$\epsilon ^{\prime}$ & 0.100 & 0.100 & 0.103 & 0.105 & 0.100 & 0.100 & 0.100 & 0.100 \\

\hline\noalign{\smallskip}
\end{tabular}
\end{table}

In Fig. 2, we present results for even-even Fm and No isotopes with
neutron number from $N=150$ to 156. This is the mass region where
active structure study is currently going on. In Fig. 2, theoretical
results for both $g$ and $\gamma$ bands are predicted up to $I=10$.
The axial and triaxial deformation parameters used in the present
calculations are listed in Table I. We note that deformation in
nuclei is a model-dependent concept. Our deformations are input
parameters for the deformed basis, and in principle, it is not
required that the numbers in Table I are exactly the same as nuclear
deformations suggested by other models. Nevertheless, it turns out
that our employed axial deformation parameters are very close to
those calculated in Refs. \cite{Cwiok94,Moller95}, and follow the
same variation trend along an isotopic chain as predicted by other
models (for example, the most deformed isotope has the neutron
number 152 and a decreasing trend for heavier isotopes is expected).
A strict test for using these parameters will be whether the
calculation can describe all observables, the most relevant one
being the $B(E2)$ value (see discussions below). The triaxial
parameter $\epsilon ^{\prime}=0.105$ that gives the correct position
of the experimentally observed excited band in $^{256}$Fm
\cite{Fm256} corresponds to $\gamma=25^o$ in terms of the usual
gamma parameter, if one uses as a rough estimate $\gamma =
\tan^{-1}(\epsilon ^{\prime} / \epsilon )$. Note that there are no
existing data that are firmly assigned to be $\gamma$-vibrational
states. The excited band starting from $I^\pi=2^+$ in $^{256}$Fm
\cite{Fm256} was {\it assumed} by Hall {\it et al.} to be
$\gamma$-vibrational states. Now our results strongly support the
interpretation of Hall's $2^+$ band as $\gamma$-vibrational band.
For those nuclei that have no $\gamma$ band data to compare with, we
simply employ $\epsilon ^{\prime}=0.1$ in the calculation. Overall,
we predict $\gamma$-vibrational bands in these nuclei, with the
rotational behavior very similar to that of $g$ bands. With the
present set of parameters, we find that the $I_K=2_2$ bandhead
energy lies low, generally at $0.6 - 0.8$ MeV above the ground
state.

The type of $\alpha$-decay experiment \cite{Ds270} makes the
identification of excited states in the heaviest nuclei possible. We
take an example from the heaviest isotopes and perform calculations
for $^{270}$Ds. With the same interaction strengths as used for the
Fm and No isotopes and deformation parameters $\epsilon=0.225$ and
$\epsilon^\prime=0.1$, we obtain the sequence of ground state
rotational band energy $E(I_K=2_0)=0.048$, $E(4_0)=0.159$,
$E(6_0)=0.332$, $E(8_0)=0.568$, and $E(10_0)=0.863$ (all in MeV), in
a good agreement with the experimental estimation \cite{Ds270}.
Furthermore, we predict very low-lying collective $\gamma$
vibrations, with the bandhead energy $E(2_2)=0.565$ MeV. The results
will be presented later in the discussion on $\gamma$ phonon states.
We note that at present, there is no better way to determine the
triaxiality for $^{270}$Ds, and using $\epsilon^\prime=0.1$ for the
whole mass region is a natural choice. What we want to emphasize is
that with this assumption, there is a low-lying $\gamma$ band in
this truly superheavy nucleus.

Recently, a great deal of experimental effort has been made to
understand low-lying excited structure in transfermium nuclei. The
discussion so far has been focused on quasi-particle excitations
only \cite{Rodi07,Rodi06,Tandel06,Green07}. Our present results
suggest that there must exist an important part of low-energy
collective vibrations, in particular at the energy region below 1
MeV. It is easy to make the following estimation. The pairing energy
gaps from our BCS calculation for $^{256}$Fm are $\Delta\approx 0.5$
MeV for neutrons and $\Delta\approx 0.8$ MeV for protons, which are
needed amount to correctly reproduce the rotational behavior (e.g.
moment of inertia). Thus, energy of a 2-quasiparticle state should
be greater than 2$\Delta$, i.e. 1 MeV. Therefore, we expect that a
2-quasiparticle excitation energy is generally higher, which lies at
or above 1 MeV in this mass region. This is a useful criterion to
identify a low-lying excited state as a collective state rather than
a quasiparticle excitation. Of course, certain residual interactions
may push the quasiparticle configurations down. Very recently,
possible occurrence of low-lying alternative parity bands in
superheavy nuclei due to octupole correlations has been suggested
\cite{Shnei06}.

\begin{figure}
\includegraphics[width=5.5cm]{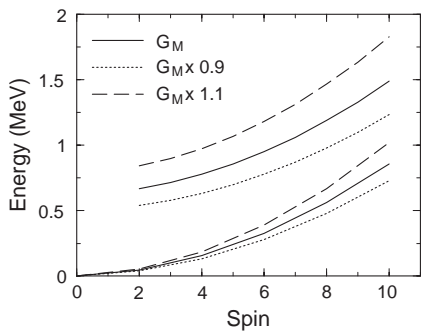}
\caption{Comparison of calculated $g$ band (lower group of curves)
and $\gamma$ band (upper group of curves) for three sets of pairing
parameters: the original pairing (solid curves, which reproduce the
data, see Fig. 2), the one multiplied by a 0.9 factor (dotted
curves), and the one multiplied by a 1.1 factor. $^{256}$Fm is taken
as the example.} \label{fig3}
\end{figure}

A detailed prediction for $\gamma$ band depends on pairing strengths
which are parameters in the model. The parameters $G_1$ and $G_2$ in
(\ref{Pairing}) are adjusted in accordance with the size of
single-particle space, and are chosen so that they can give correct
rotational sequences for the superheavy mass region (as presented in
Fig. 2). Figure 3 shows the deviation from the current prediction if
the pairing strengths $G_M$, and thus $G_Q$ (since $G_Q$ is
proportional to $G_M$), are allowed to vary by $\pm 10\%$. It is
observed that the curvature of the curves, i.e. the rotational
frequency $\omega=\Delta E/\Delta I$, increases with increasing
pairing strength. Pairing also shifts the $\gamma$ bandhead; a
stronger pairing leads to a higher bandhead. In Fig. 3, about 20$\%$
of deviation is seen for the $\gamma$ bandhead when pairing changes
by 10$\%$. However, even with this amount of uncertainty in pairing,
the predicted $\gamma$ bandhead energy is still within 1 MeV of
excitation, and thus the conclusion for the occurrence of low-energy
$\gamma$ vibrations remains valid.

\begin{figure}
\includegraphics[width=8.5cm]{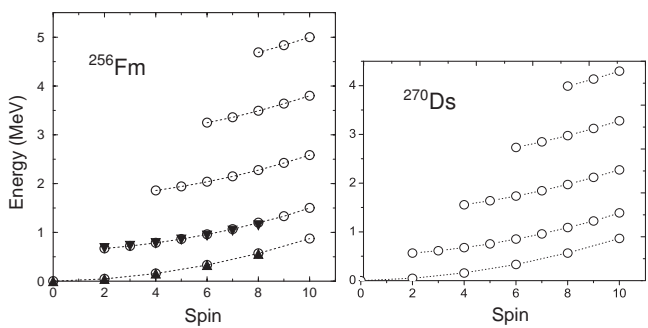}
\caption{Predicted multi-phonon $\gamma$-vibrational spectra in
$^{256}$Fm and $^{270}$Ds. Available experimental data of $^{256}$Fm
\protect\cite{Fm256} for the $g$ band and $\gamma$ band are also
shown.} \label{fig4}
\end{figure}

\begin{table*}
\caption{Calculated in-band $B(E2, I_0\rightarrow (I-2)_0)$ values
for $g$ bands, inter-band transitions from $\gamma$ bands to $g$
bands $B(E2, I_2\rightarrow I^\prime_0)$, and from 2$\gamma$ bands
to $\gamma$ bands $B(E2, I_4\rightarrow I^\prime_2)$. Numbers are
given in W.u..}
\label{tab:2}
\begin{tabular}{cccccccccc}
\hline\noalign{\smallskip}
$I_{K,i}\rightarrow I^\prime_{K^\prime,f}$ & $^{250}$Fm & $^{252}$Fm & $^{254}$Fm & $^{256}$Fm & $^{252}$No & $^{254}$No & $^{256}$No & $^{258}$No & $^{270}$Ds \\
\hline\noalign{\smallskip}

$2_0\rightarrow 0_0$ & 314 & 341 & 333 & 297 & 316 & 351 & 334 & 301 & 282 \\
$4_0\rightarrow 2_0$ & 449 & 488 & 476 & 425 & 452 & 501 & 478 & 431 & 404 \\
$6_0\rightarrow 4_0$ & 496 & 539 & 526 & 470 & 500 & 554 & 528 & 476 & 448 \\
$8_0\rightarrow 6_0$ & 522 & 567 & 554 & 496 & 526 & 582 & 555 & 502 & 473 \\
\hline\noalign{\smallskip}
$2_2\rightarrow 0_0$ & 3.45 & 3.61 & 3.72 & 3.97 & 3.68 & 3.60 & 3.57 & 3.64 & 4.19 \\
$2_2\rightarrow 2_0$ & 5.18 & 5.47 & 5.63 & 5.83 & 5.51 & 5.45 & 5.30 & 5.27 & 6.60 \\
$2_2\rightarrow 4_0$ & 0.29 & 0.31 & 0.32 & 0.31 & 0.30 & 0.31 & 0.29 & 0.27 & 0.40 \\
$3_2\rightarrow 2_0$ & 6.16 & 6.45 & 6.64 & 7.09 & 6.57 & 6.43 & 6.37 & 6.49 & 7.48 \\
$3_2\rightarrow 4_0$ & 2.75 & 2.95 & 3.02 & 3.02 & 2.92 & 2.93 & 2.79 & 2.68 & 3.72 \\
$4_2\rightarrow 2_0$ & 1.89 & 1.96 & 2.02 & 2.23 & 2.03 & 1.96 & 1.98 & 2.08 & 2.13 \\
$4_2\rightarrow 4_0$ & 6.35 & 6.72 & 6.91 & 7.17 & 6.76 & 6.69 & 6.51 & 6.48 & 8.11 \\
$4_2\rightarrow 6_0$ & 0.64 & 0.71 & 0.73 & 0.70 & 0.68 & 0.71 & 0.65 & 0.60 & 0.96 \\
$5_2\rightarrow 4_0$ & 5.27 & 5.48 & 5.66 & 6.14 & 5.63 & 5.48 & 5.48 & 5.67 & 6.17 \\
$5_2\rightarrow 6_0$ & 3.59 & 3.91 & 3.99 & 3.94 & 3.82 & 3.87 & 3.66 & 3.48 & 5.03 \\
\hline\noalign{\smallskip}
$4_4\rightarrow 2_2$ & 15.3 & 15.6 & 16.0 & 16.7 & 15.8 & 15.5 & 15.6 & 15.7 & 17.3 \\
$4_4\rightarrow 3_2$ & 9.04 & 9.21 & 9.38 & 9.59 & 9.31 & 9.09 & 9.04 & 8.99 & 10.3 \\
$4_4\rightarrow 4_2$ & 3.39 & 3.43 & 3.46 & 3.46 & 3.46 & 3.37 & 3.31 & 3.23 & 3.85 \\
$4_4\rightarrow 5_2$ & 0.74 & 0.75 & 0.75 & 0.72 & 0.75 & 0.73 & 0.71 & 0.67 & 0.84 \\
$4_4\rightarrow 6_2$ & 0.07 & 0.07 & 0.07 & 0.07 & 0.07 & 0.07 & 0.07 & 0.06 & 0.08 \\
$5_4\rightarrow 3_2$ & 10.1 & 10.4 & 10.7 & 11.2 & 10.5 & 10.3 & 10.4 & 10.6 & 11.4 \\
$5_4\rightarrow 4_2$ & 10.9 & 11.1 & 11.4 & 11.7 & 11.2 & 11.0 & 11.0 & 11.0 & 12.4 \\
$5_4\rightarrow 5_2$ & 5.72 & 5.80 & 5.86 & 5.86 & 5.84 & 5.71 & 5.60 & 5.46 & 6.51 \\
$5_4\rightarrow 6_2$ & 1.58 & 1.60 & 1.59 & 1.54 & 1.60 & 1.57 & 1.51 & 1.43 & 1.81 \\
$5_4\rightarrow 7_2$ & 0.18 & 0.18 & 0.18 & 0.17 & 0.18 & 0.18 & 0.17 & 0.16 & 0.21 \\

\hline\noalign{\smallskip}
\end{tabular}
\end{table*}

Multi-phonon $\gamma$ bands are rotational bands built on top of
$\gamma$ vibration classified by phonons. In Fig. 4, we plot all the
states for spins $I\le 10$ obtained by diagonalization for
$^{256}$Fm and $^{270}$Ds. One sees that the excited states are
clearly grouped according to $K=0,2,4, \dots$. We identify the first
excited band ($K=2$) as the $\gamma$ band, the second excited band
($K=4$) as the 2$\gamma$ phonon band, the third excited band ($K=6$)
as the 3$\gamma$ phonon band, and so on, with presence of strong
anharmonicity in vibration. We have calculated all eight nuclei
discussed in Fig. 2, and obtained similar patterns as for
$^{256}$Fm. In particular, we predict $2\gamma$ phonon bands for
them with the bandhead energy at about 1.8 MeV. $2\gamma$ phonon
states were observed in the rare earth nuclei $^{166}$Er and
$^{168}$Er. To compare with those $2\gamma$ phonon bandhead
energies, 2.029 MeV in $^{166}$Er \cite{Er166} and 2.056 MeV in
$^{168}$Er \cite{Er168}, $2\gamma$ phonon states in superheavy
nuclei lie lower. No 3$\gamma$ phonon states have yet been seen
experimentally in any known examples of nuclear spectroscopy.
According to our calculations, they should appear at about 3 MeV in
Fm and No isotopes. To compare with $^{256}$Fm, a more compressed
multi-phonon $\gamma$ vibrational spectrum is seen for $^{270}$Ds,
again with strong anharmonicity.

The wave functions obtained after diagonalization of the Hamiltonian
are used to calculate the electric quadrupole transition
probabilities
\[
B(E2: (I_i, K_i)\rightarrow (I_f, K_f))=\frac 1{2I_i+1}\left|
\left\langle \Psi_{I_f,K_f} || \hat Q_2 || \Psi_{I_i,K_i}
\right\rangle \right| ^2
\]
between an initial state $(I_i, K_i)$ and a final states $(I_f,
K_f)$. The explicit expression for the reduced matrix element in the
projected basis can be found in Ref. \cite{Sheikh01}. Note that we
now use $K$ instead of $\sigma$ to specify states with the same
angular momentum $I$ to keep the familiar convention. In the
calculation, we use the standard effective charges of 1.5$e$ for
protons and 0.5$e$ for neutrons.

In Table II, we list calculated $B(E2)$ values within the $g$
bands, the inter-band linking transitions between $\gamma$ bands
and $g$ bands, and between 2$\gamma$ bands and $\gamma$ bands. It
is found that for each of the nuclei, the $g$ band $B(E2)$ values
correspond to a rather constant transition quadrupole moment,
reflecting the fact that these systems are good rotors. The
inter-band transitions are on average by two to three orders of
magnitude smaller than the in-band transitions within the $g$
bands. These predicted $B(E2)$ values may help in determining the
structure of low-lying states.

To summarize, we have applied the Triaxial Projected Shell Model to
Fm and No isotopes as well as $^{270}$Ds to predict $\gamma$
vibrational states in superheavy nuclei. Shell model diagonalization
is carried out in a angular-momentum-projected triaxially-deformed
basis. It is found that the calculation simultaneously leads to a
consistent description of ground state band and multi-phonon
$\gamma$ bands in these nuclei. The physics of the bands is
discussed in terms of $K$ mixing, suggesting a microscopic
connection between the excited vibrational states and the nuclear
ground state properties in the heaviest nuclei where structure study
can be performed experimentally. In- and inter-band $B(E2)$ values
are predicted. This work calls for attention on collective
excitations in the low-energy region where active structure studies
are currently carried out.

Communication with R.-D. Herzberg, P. M. Walker, and F.-R. Xu is
acknowledged. Y.S. thanks the colleagues at the Institute of Modern
Physics, Tsinghua University, and Peking University, China, and at
University of Surrey and University of Liverpool, U.K., for the warm
hospitality extended to him. This work is supported in part by the
Chinese Major State Basic Research Development Program through grant
2007CB815005, the National Natural Science Foundation of China under
contract No. 10325521, and the U. S. National Science Foundation
through grant PHY-0216783.



\end{document}